# Randomization to Reduce Terror Threats at Large Venues

Paul B. Kantor, Fred S. Roberts

Rutgers University 96 Frelinghuysen Road, Piscataway, NJ 08854-8018

paul.kantor@rutgers.edu; froberts@dimacs.rutgers.edu

## 1. Executive Summary

Can randomness be better than scheduled practices, for securing an event at a large venue such as a stadium or entertainment arena? Perhaps surprisingly, from several perspectives the answer is "yes." This note examines findings from an extensive study of the problem, including interviews and a survey of selected venue security directors. That research indicates that: randomness has several goals; many security directors recognize its potential; but very few have used it much, if at all. Some fear they will not be able to defend using random methods if an adversary does slip through security. Others are concerned that staff may not be able to perform effectively. We discuss ways in which it appears that randomness can improve effectiveness, ways it can be effectively justified to those who must approve security processes, and some potential research or regulatory advances.

## 2. Introduction

When we face an intelligent adversary, sometimes randomness is absolutely required. The familiar game of "rock, scissors, paper" shows this. Each alternative is beaten by one of the others, and beats the remaining one. Thus, paper *beats* rock *beats* scissors *beats* paper. If we play any fixed schedule of choices, our adversaries, once they learn the pattern, can always play the better choice. So, the best we can make is to play each alternative, randomly, aiming to do so about one third of the time[1]. We can do this by rolling a die, and playing rock for 1 or 2, scissors for 3 or 4, and paper for 5 or 6. This is an example of a powerful general theorem from the field called Game Theory, showing that randomness can be better than any fixed schedule.

This note builds on key findings of a study on applying randomness to security at large venues, conducted by the CCICADA Center at Rutgers University. The findings are in three parts: goals of randomization, that is, ways that venues have considered to justify its use; methods of randomization, that is, practices that achieve some degree of randomness; and, finally, thoughts about future research or regulatory changes, to improve security.

## 3. Methods

The project team conducted extensive literature review, conversations with experts on venue security, and a survey of selected leaders in the field. They also consulted some experts in other fields, such as port security or airport security, looking for best practices not yet used at venues.

---

[1] If we play any one alternative, say "paper" more than any of the others, the adversary should play to beat it ("scissors") *all the time*. He or she will beat us more than we beat him or her.

They often found widespread expectation that a procedure is desirable, with little concrete or objective supporting evidence, and with very little uptake at venues.

The project included an analysis of statistical data from the survey. To do this they boiled down scores of practices found in the literature or conversations to some 25 specific random practices. An example is: "*S1 At a normal bag check table randomly select some bags for an added check, using an available technology (explosives screening; liquid scanner; x-ray; canine)*"[2]. Table 1 shows the instructions to the experts.

*Table 1. Instructions of the Expert Assessment of Selected Practices*

| |
|---|
| It is becoming known that if you already do some security practice A (at 100%), adding another security practice B at a 30% level has very little value if it is predictable/avoidable. But if it is done randomly, it adds about 30% of whatever benefits practice B offers, and it increases the deterrent effect.<br>The following examples of random practices are suggested by our interviews, a review of the literature and our research and observations. Please score them for us. There is space to add any of your own suggestions at the bottom of the list. You may leave this sheet and come back to continue it, until you finally put an X in the "submit" box at the very bottom. |
| We'd like you to score them on all these criteria<br>    Experience:    Have you actually used it (not necessarily at your present venue)?<br>    Importance:    An important practice should be part of any venue's approach to security, and adding it will significantly strengthen security<br>    Feasibility:    A practice is feasible if it can be added to your present policies without major costs and personnel challenges<br>    Sustainability:   After a successful and effective launch, a sustainable security approach/measure will not decay in its implementation, while it remains relevant, due to various factors such as cost, physical/stamina capabilities of staff, stadium or crowd dynamics, equipment wear-and-tear or failure, follow-up auditing, complacency, etc. |
| This survey has four parts:<br>    (1) Assess specific practices;<br>    (2) Assess ways of achieving randomness;<br>    (3) Combining the two;<br>    (4) About the balance that randomness should play in a security plan.<br>You may go back and adjust your answers if you like. We have designed it so that part (3) appears only after you have worked through parts (1) and (2), and select the option to do part (3). |

---

[2] Like a number of suggestions, this one was scored as important by more than 50% of the experts; also a majority considered it feasible, and a majority judged it sustainable as well. Yet only 17% of them had any experience actually doing it.

The detailed practices and survey findings were provided by CCICADA to the DHS Office of SAFETY Act Implementation (OSAI).

Survey response shows that: importance, feasibility, and sustainability together do not ensure that a practice *has actually been used.* For many specific practices, concerns about staff capability, training, or the need for a cost justification, prevent or delay implementation. Still, high scores on importance, feasibility and sustainability suggest that the practice could be considered a "best practice," until evidence from the field may provide a more objective assessment of its value.

**Goals of Randomization**

The survey also asked what goals experts hope to serve with randomness. As for the practices, the project assembled a large set of possible goals, and narrowed them down to four that were included in the survey. Table 2 shows these key goals for randomization, together with a note about how prevalent each goal is, among the experts who were consulted.

*Table 2 Goals Associated With Randomization*

| Goal of Randomization | Details | Evidence |
|---|---|---|
| (1) Deceiving or confusing an adversary; | When there is an unknown chance of being intercepted, the cost of each penetration becomes higher, and the perceived self-efficacy (chance of succeeding) goes down | Mentioned by essentially all experts consulted |
| (2) Monitoring operational effectiveness; | An example is random rechecking the credentials of contracted workers who are initially checked by the contracted service | Mentioned by some security experts |
| (3) Stimulating the capability and/or alertness of security staff personnel; and | The clearest example of this is random use of "red-teamers" who attempt to smuggle contraband past screeners. There is a competition between "stay at the same job, and do it efficiently" and "become careless because of the routine" | Mentioned by few security experts. Several kinds of devices can be programmed to issue random alarms, which can be used, among other things, to assess alertness |
| (4) Achieving intermediate levels of security when threat intelligence and/or budget considerations do not match well to any specific security posture. | If some fraction of cases (persons; CCTV cameras) can be examined more thoroughly, without exceeding available resources, that is better than not doing it. Randomness confounds an adversary | Not mentioned by many security practitioners. Mentioned by engineers. |

As Table 2 shows, venue security directors are generally aware of the first two aspects, with some awareness of the third. The fourth, which seems well known in some other technical areas (such as missile defense) is known to engineers involved in developing or testing security devices, but is not yet widely recognized among expert practitioners in venue security.

These goals are not "orthogonal." Any particular practice may advance one, two, or even all four of these goals. However, all security operations, including the use of randomization, are conducted in the context of overarching (organizational) goals. Those overarching goals cannot be ignored when focusing on randomization. They are: (1) providing adequate security for all persons and facilities at every kind of event; (2) enhancing profitability of the venue, over the range of its operations; (3) providing the best possible "patron experience" at each of the events.

The overarching goals are interrelated: security contributes (by preventing attacks) to increased revenue; good patron experience also contributes to revenue; and increased revenue increases the range and intensity of possible security practices. On the other hand, there are innate tensions among the three overall goals. Every dollar applied to security is not available as profit; rigorous security measures may diminish the patron experience; and so forth.

The experts who were consulted provided many insights when asked why they did (or did not) try, or continue, a specific practice. Some of these insights appear as text boxes in this note, giving the experts "their own voice" about the problem.

**Methods of Randomization**

The study found that "randomization" has several definitions.

> *We counted out every so many people and the supervisor would put a light on in what was supposed to be random lanes for additional screening … a red light that the screener could see, [it].was supposed to be random, but more often turned into [doing the lanes in turn]*
>
> *Security director*

Table 3 shows the most important of these interpretations. All are clearly different from "determined and known." They represent various ways of opposing an adversary who seeks to get around the procedures. The most common term used is "unpredictable," which is, after all, a good cognate for randomness. Several of these ideas exhibit procedural or legal weaknesses. Many experts are aware of these weaknesses, and some common themes are shown in quotation boxes.

Of the seven definitions extracted from literature and interviews, the project retained the first five in its survey, inviting respondents to present other methods, based on their own experience

> *[We tried "every K-th person, our] limited experience seemed to result in unintentional profiling*
>
> *Security director*

*Table 3 Selected Ways to Interpret Randomization*

| Practice | Detail | Evidence |
|---|---|---|
| R1 every *n* persons[3] | Variants of this method are reported at many venues as a kind of "randomization" | Interviews. The spacing ("*n*") may change based on demographic and threat level |
| R2 visible random device | Examples include a touch device that patrons activate, which may signal additional screening | Discussions with security personnel; analysis of perceptions |
| R3 hidden random device | Examples are a photocell or other counter on a WTMD, which issues random alarms | Many commercial devices have this capability; it may be used to invoke "random added screening" |
| R4 today "do or not" | Example might be the use of added security canines; the decision would be unpredictable | Conversations with security personnel; the decision is most often coupled with demographic or threat information, and/or resource availability |
| R5 random plan by last minute distribution of one from a set of prepared plans of procedure (Playbook) | The plan might detail which employees are assigned to specific tasks, etc. | Conversation with TSA expert. Such methods are constrained by employee capabilities, but do control insider threats |
| R6 spur of the moment choice by security personnel | A security director may visit the broadcast trailer on an unpredictable schedule | Expert interviews |
| R7 spur of the moment change by an individual employee | This is most likely to occur when crowd pressure, or an intuitive judgment leads to skipping a security process | Observations of screening processes. This is not recommended. Even spur-of-the-moment *added* screening may reflect or be perceived as reflecting bias. |

Practical considerations always affect implementation of randomized procedures. A bomb-sniffing canine cannot move instantly from the East stands to the West stands; randomization must let both the dog and the handler move along a reasonable (otherwise random) path.

> *While we don't keep a specific list we do alter sweep routes ...we do this regularly*
> 
> *Security director*

---

[3] Because it is not truly random, this method could be defeated by a three-person team. The advance member clears security and stands with her cell-phone, counting to determine the spacing of the chosen persons. The other two switch places if needed, as they enter screening, to arrange that the one carrying the weapon is not the n-th person.

Discussing randomization often leads to other practices that do not affect all patrons equally. Some experts mentioned the use of behavioral information ("behavioral profiling") to select patrons for more intense scrutiny. For example:

> *Senior security members routinely go outside and do random perimeter checks*
> *Security director*

"We noticed that one woman was wearing a winter coat in July. So we asked her about it. She explained that she did it because of a skin condition, and we were satisfied." Another "non-uniform" policy is to strengthen security procedures when there is non-public information about an increase in risk.

Neither of these examples is really "random." Apparently they are mentioned because experts believe that adversaries cannot know when the indicated challenge or change will happen. That makes them "unpredictable, from the perspective of the adversary," and so they might be considered "random." In contrast, "tighter security shortly after a widely publicized attack," would not seem to be "random."

*Deceiving or confusing an adversary.*

Deception is the goal that was most frequently mentioned by experts. Paraphrasing one example: "if someone is watching us through binoculars, and sees that after the dog checks a trash can for explosives, the dog sometimes comes back later, he cannot tell when it is safe to put his bomb in that can." Or, as another expert put it, somewhat more colorfully: "We don't want to be like the German camp guards in 'Hogan's Heroes' who guarantee some 19 minutes of safe time because they never vary their schedules."

Several informants indicated that random assignments of personnel to tasks (such as being responsible for different parts of a stadium in the pre-game sweep) also helps control insider threats. An adversary cannot know whether an accomplice will be the one to screen a particular section, and even the insider accomplice could not know where to plant a bomb, without knowing where he or she will be screening on game day.

*Is it Detection of Deterrence?*

As soon as we consider the adversary's reasoning about any particular practice, its effect may also be characterized as deterrence. To paraphrase one respondent, "when they realize they cannot figure out what I'm doing, they'll move on and look for an easier target."[4]

---

[4] This observation highlights the fact that security directors' primary responsibility is to their own patrons, though it is not to suggest that they don't care about attacks on others' venues, which in turn would have considerable indirect impact on their own.

## 4. Strengths and Limitations of Randomness

**Randomness *versus* Completeness**

As noted in Table 1, randomness can be used in extending and strengthening a security program. But pure randomness also creates gaps. An interesting example comes from policies for randomly repeating background checks on employees. One recommendation, from a Federal agency, is that every employee have a 10% chance of being rechecked every year, *even if he or she was re-checked last year, and/or three years ago, etc.* The rule is "every year one tenth of all employees are selected *randomly* and re-checked." Now, it might seem that if one tenth are checked every year, in ten years, all of them will have been checked, and, in fact that same agency makes this inference.

> *Another randomized practice was the scanning of credentials on the field and in the Clubhouses. This practice allowed security/League staff to validate credentials on anyone that might look "out of place" or had a fraudulent credential.*
> *Security Director*

But the very randomness works against us here. Each employee has a 90% *chance of not being picked*, every year. Suppose there are 1000 employees. Then 900 of them are not picked in the first year. 90% of these, or 810, are also not picked in the second year. If we continue this computation of 90% not picked, after 10 years about 35% will still have not been picked and re-checked.[5]

For the 1000 employees this means that about 349 of them, or more than one third (!) will not be re-checked by this random process. Since randomness means that some people are re-screened two or even more times, there is no way to avoid missing some of the others[6].

> *[For red-teaming] We hired college aged [workers] to probe and gave them instructions to be creative, they didn't do the same location twice and if they did, they tried a different tactic with a different actor.*
> *Security Director*

**Monitoring operational effectiveness**

The random process most used at venues is random use of "test subjects" with known threats. Most commonly this is "red teaming" or "covert testing" in which an accomplice with a weapon or other contraband attempts to pass security.[7] This method is considered random, insofar as the scheduling of the test, by time and location, is intended to be hard to predict. One expert suggested that in the course of screening for a game, the covert red teamer would test or probe every one of the security screening stations. Of course this is not fully random, but shows an interesting balance of completeness and unpredictability.

---

[5] Things are a little better if we check half of them each year (much more expensive!). Even so, after two years fully 25% of them will remain unchecked.

[6] This has been discussed by academic researchers. See "To Read More" below.

[7] This method was applied with alarming results in a GAO study of practices at border points of entry. See: BORDER SECURITY: Summary of Covert Tests and Security Assessments for the Senate Committee on Finance, 2003–2007 GAO-08-757 Available at: https://www.gao.gov/assets/280/275595.pdf

Red teaming directly measures detection rates, but does not give any indication of how many threats there are. Since detection is surely less than perfect, randomly *re-screening* some patrons who have *already passed screening* may measure the detection level and also give an indication of threat prevalence. For example, random re-inspection can be used at heavily trafficked border stations (the COMPEX program), to estimate the "fraction of threats that are missed." This requires very heavy traffic and may not work at venues. For example, if detection is 70% effective, some 30% of all contraband will be missed. Suppose we use a small random sample, say 10% of the cleared patrons, to estimate misses and prevalence. If contraband is not very common (say twenty or thirty cases per event), then doing this for a whole event with very thorough searches of the random selectees, would find 100% (detection rate at the second screen) of 30% (previously missed) of the 10% carried by people selected for random screening. This is just 3% of, say, 30 items. That is barely one item per event. Such a tiny number cannot support a precise estimate of the detection rate or the true prevalence of contraband.

**Stimulating the capability and/or alertness of security staff personnel**

Some experts report that having supervisory personnel follow unpredictable schedules increases the awareness and alertness of line personnel. To paraphrase one, "the media people like to think that they handle their own security, but I want them to know that I might drop into their trailer at any time, and see for myself if they are doing what they say."

Other examples, as mentioned above, are the benefits of random or unpredictable red teaming in motivating alertness. Some venues provide modest rewards for staff who detect such threats, while others report that "word gets around if someone misses the mark, and [then] everyone else steps up their game."

**Achieving intermediate levels of security by random combination of methods**

This strategy (for example, as in (Kantor et al. 2008, 2010)) is well known to engineers studying sensor optimization but is not widely recognized in the security community. Conversations with engineers involved in designing or testing security devices show that they are aware of its potential for improving security. However, the study found that venue security practitioners generally don't know much about this technical idea. This represents an important possibility for increasing security, discussed in Section 6 .

Currently some experts consider a partial mixing to increase security. Suppose that a venue can afford to do "a little more screening" of arrivals than it does. But it cannot afford to do a higher level of screening for all of them. It makes sense to do it for a fraction of the arrivals, and, in line with goal (1) of Table 1-1 (deceiving or confusing an adversary), it is best to select the subjects for stronger screening randomly. When some experts use a phrase like "randomization can increase security" they may refer to the added detection capability of random combinations, or to the deterrent effect of using randomization. Deterrence does not change detection rates (there is actually less to be detected). But it does increase overall security.

> *Our number one random method currently is the use of area CBRNE monitors that alert to chemical, biological, radiological substance that requires additional vetting, along the same line, use of Vapor wake explosive detection canines*
>
> *Security Director*

## 5. Benefits and Objections for Random Methods

For some randomization approaches in Table 3 the study found frank unwillingness to adopt them. For others, while widely mentioned, they are sometimes modified in ways that make them less random.

Methods Included in the Survey

**Choosing "every n$^{th}$ person"**
This approach is often mentioned, with numbers such as seven or ten being common. It is more common for it to be used for additional scrutiny, than for reduced scrutiny. There is not much resistance to this, and it has proven easy to use, and easy to explain to those patrons selected for extra screening. In practice, when the selectee appears to be "obviously not a threat," front line personnel may, however, "bend the rule" and skip to the next person. This diminishes randomness.

> *. Randomization can be risky in the fact that it takes away from that individuals focus on their task at hand. If randomization is implemented in such a way that it's fluid (i.e.: managed by the supervisor, automated, etc.) that might be the most effective.*
>
> *Security Director*

**Selection using a visible random device**

There are reports about using some device, such as a floor mat, connected to a random generator. The patron activates the device by stepping on the mat, and a random number shows up. This gives the process some "visible patron agency." This might also be done with a random number generator on a tablet or smartphone, and a posted "number of the day" that all patrons can see. Patrons would see those ahead of them activate the device, and see that most are not selected. This enhances the sense of control, and perception of randomness. While there do not seem to be objections, none of the experts consulted had actually used this kind of method.

> *"Always the same"" provides a more structured and consistent program for the staff. .... streamlined training and accountability. A more randomized process helps deter those who are looking for patterns; also helps keep our staff more alert and aware. There is room for both in a good operation.*

**Selection using a hidden random device**

A walkthrough detector might include a photocell that counts patrons. It might be coupled to a random selector, set to display a random integer. Selection of a number with positive connotations, such as 3 or 7, may have some psychological advantages, enabling security personnel to say something like "you are a lucky three" to a person selected at random.

> *Although we don't control parking lots we do have vehicles on perimeter that are screened. K-9 does do random checks as well*
>
> *Security Director*

> *Limited experience [with randomization] seemed to result in unintentional profiling*
>
> *Security Director*

**Decide, on the day, whether or not to do a particular activity**

This approach represents a plausible comprise between strict randomness and the requirements imposed by time and space, personnel training, and other resource limitations. If there is a fixed overall schedule, then it is susceptible to insider threats. Unless the selection process itself includes true randomization this approach may become so routine that the overall policy becomes predictable.

**Last minute distribution of a random selection from a prepared set of plans**

Plans might detail which employees are assigned to specific tasks, etc. This approach remedies the weakness of patterned selection. Generally, the larger the set from which one plan is chosen, the more surveillance an adversary must do to learn the entire array of plans. And, even if that happens, the adversary will not know which plan is to be used on a given day. Similarly, insiders will not know until their shifts begin what plan is in effect today. This was mentioned both by a TSA expert and by one venue security director. Use of such methods does help control insider threats, but it is constrained by employee capabilities.

Methods not included in the Survey

Some methods of randomization that arose in the literature or in interviews were not included in the survey, and we mention them here for completeness. The first two are not sound practices. The remaining two seem to require sophistication and cross-training of employees, and may not be practical in today's venue security setting.

> *Due to manpower, [we] just asked LEO to move around, and not s[t]ay in anyplace more than 5 minutes, not really mathematically random, but they moved not in a linear fashion and skipped around to make their presence more robust by being seen on different sides of the venue within minutes.*
>
> *Security Director*

**Spur of the moment choice policy for front line security personnel**

While this might sound plausible, multiple respondents observed that people doing a heavily routine job are quite likely to also adopt routines when asked to be spontaneous. Thus, it is risky to think that asking for this kind of "spur of the moment choice" will work. As one respondent put it, "if Joe usually turns to the left first, and then to the right, that's not random." In patron screening this type of choice is also most likely to occur when crowd pressure, or an intuitive judgment leads to *skipping* a security process. Since it is triggered by perceived crowd pressure, it is not unpredictable. Even spur-of-the-moment *added* screening may create some problems either if it reflects employee biases, or if patrons perceive it as biased.

**Spur of the moment change by an individual employee**

> *The front line of defense, must see in black and white, while the supervision must see in the overwhelming gray areas and react accordingly.*
>
> *Security Director*

In this approach an employee might change procedures based on some kind of internal decision process. This is limited by employee skill and training constraints. As one person interviewed noted: "to do a routine job effectively you have to fall into a pattern. If you have to keep switching it's easier to make a mistake."

**Using (pseudo-) random[8] data to select an action, at regular times**

Here an employee would change, for example, every half hour, from one activity to another. This could be as simple as: "every half hour toss a coin. If it's heads, continue on clockwise; if it's tails, retrace your path." With this, as with other random patterns, there is a concern that the method will leave some important part of the overall security plan uncovered.

---

[8] The modifier "pseudo" recognizes that numbers generated by computer algorithms are not technically random. But since it can take hundreds of years for them to repeat, we can say that (if the key or "seed" of the algorithm is kept private) they are random for all practical purposes.

**(Pseudo-) random timing of an action or one from a set**

If the timing of the change is unpredictable things are certainly harder for an adversary. But, perhaps even more than other methods of being random, this one is limited by employee skills. On the positive side, it might offer a motivational advantage, as work is definitely less routine. However, whether this is a benefit or a weakness seems to depend on both the task and employee. From the management perspective, it is very difficult to schedule employees to do something "at random times" without incurring substantial down time.

## 6. Randomization to Increase Security

The survey did include the following possibility for randomly changing the security or screening setting.

*S4 Instead of having all WTMDs set at the same level the whole time for a specific event, use some kind of randomization (by time, by lanes, etc. This is easier with networked WTMDs).*

This option was motivated by study of the engineering literature on sensors and threat detection. In engineering multiple techniques can be combined in an approach called "data fusion for signal detection" (Chair and Varshney, 1986). An application to cargo screening is given by (Boros et al., 2010; Kantor et al., 2008; Kantor and Boros, 2010). However, only one among the expert respondents has ever even tried this. This alternative drew comments such as "*do not see the benefit,*" "*might raise liability issues.*" "*we alter settings based on risk assessment and type of event,*" and "*not sure I would prefer randomizing to screening all.*" Fewer of the responding experts had used this than any other practice. However, this option, and others like it, may open a path towards low-cost improvement in security, with some important advantages in practice. To clarify this, let us work out a specific, if hypothetical, example.

> *…..it is dangerous to allow the same guard to be random in his/her duties as it creates room for inefficiency*
>
> *Security Director*

Suppose that a venue has the option of two practices, denoted by A and B, for screening all visitors. See Table 4. One of them is known (based on red-teaming, or on results from government testing laboratories) to be better at detecting contraband (95% success vs. 75% success). Not surprisingly, that more sensitive method (A) also takes somewhat longer to screen each individual patron 10 seconds vs. 6 seconds). Usually a more sensitive method has more delay because it also has a higher false alarm rate and more people require extra screening or divestment, which slows things down.

*Table 4  Randomization to Increase Security*

| Practice | Fraction Detected | Fraction Missed | Average time per patron | Details |
| --- | --- | --- | --- | --- |
| A (more sensitive) | 95% | 5% | 10 seconds | Everyone goes through a WTMD set at level 3, and there is a pat-down if the detector alarms |
| B (quicker) | 75% | 25% | 6 seconds | Everyone goes through a WTMD set at level 2 [league standard], and there is a pat-down if the detector alarms |
| 50-50 mix | 85% | 15% | 8 seconds | 50% randomly are screened with practice B, and 50% with practice A. |

Suppose that for an upcoming event, the known size of the crowd requires that the screening time be no more than 8 seconds per patron (that is, a throughput of 7.5 patrons per minute, on the average). Clearly, practice A, with its better security, will not meet this requirement. The natural (and usual) choice is to work with practice B, which, we suppose, meets the league standards.

Now consider a *mixed* alternative. Use some simple randomization scheme (in this case, it could be a coin toss) so that half the patrons are screened at level B, and half at level A. Since these two groups (after all, they are selected *randomly*) are equally likely to have contraband, the expected fraction of contraband detected will be the average of 95% and 75%, or 85%.

$$\tfrac{1}{2} * 95\% + \tfrac{1}{2} * 75\% = 85\%$$

And the average screening time will be the average of 10 seconds and 6 seconds, which is 8 seconds, exactly at the target.

$$\tfrac{1}{2} * 10\text{sec} + \tfrac{1}{2} * 6\text{sec} = 8\text{sec}$$

These results are shown in Table 4. If the average time could be set even higher (say 9 seconds), then an even larger fraction could be screened with the more secure practice A.

While the increased security of such randomized practices may not seem large, the impact is clearer if we examine what really concerns us most: the chance that a threat is *missed.* With practice B, the faster one, one in four threats (25%) is missed. With the randomized combination of practices, only 15% are missed. This is a 40% reduction (10 fewer, out of every 25 threats) in missed threats, with acceptable throughput.

Of course, it is important that the selection be random. If not, an adversary might figure out that the faster lanes are also less thorough in their screening, and always choose a faster lane.

Patrons can be randomly selected in several ways. One possibility is that a Walkthrough Metal Detector (WTMD) might have its level randomly changed from time to time. Under networked control the device might do this without either the screeners or the patrons knowing there has been a change. Otherwise, devices in adjacent lanes could be set to different levels, and patrons guided randomly to one lane or the other, in the correct proportions.

While technically attractive, this kind of random mixing of security measures poses some challenges. If patrons were free to choose among the lanes, they would likely see that the lane with Practice A has a longer queue, and would self-select into the less secure lane. Then the split would not be 50-50, nor would it be random, with the attendant risks. Methods for smoothly directing patrons must be suited to the setting. At some venues lane assignment could be far from the queues, or done where the queues are not visible – this may be possible in an indoor setting Another is to have a random directional signal, which simply sends half the people to one lane, and the rest to the other.

> *…. preferably the random decision process should be automated, not done personally by the individual security staff.*
>
> *Security Director*

For presenting the benefits of mixed screening to upper management, a chart can show the analysis of Table 4 in graphical form (see Figure 1). This shows that alternative A is over the "red line" for delay, but the 50-50 mixture, represented by the green dot, is inside the acceptable delay range, and increases detection to a better level. The vertical line here emphasizes the fact that security has improved, moving above the (notional) minimal acceptable level.

As noted, expert interviews and the survey showed that randomization's potential for a step up to higher security is not widely appreciated. Analysis of the potential for randomization to *increase* security suggests three important directions that might be explored.

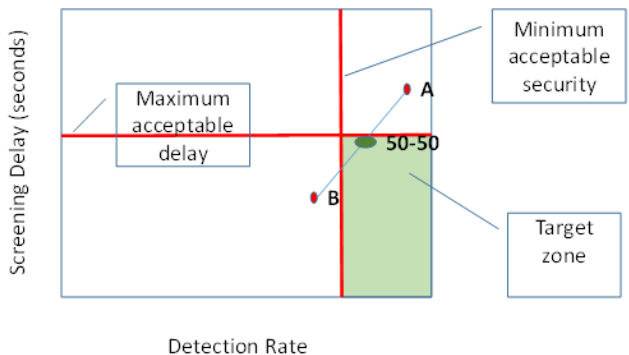

*Figure 1 Delay and Detection for a Randomized Practice*

First, the computations of Table 4 and Figure 1 assume we actually know the detection rate, and the average screening time, for both Practice A and Practice B. In reality, the venue security manager may not know any of these facts. Before these advances in security are workable: one needs ground truth for detection; ground truth for delays; and usable calculator and simulation tools.

**Ground Truth Detection Rates**. Published detection rates, established in laboratories, are not the detection rates in the field. Distractions, crowd noise, sunlight, and other factors can cause the workers monitoring screening devices to miss some (perhaps as many as 15%) of the "alarm signals." To make an honest calculation the Security Director will want to know the real rates, when done by his or her actual screening personnel. This could be done by establishing data collection guidelines so that individual venues can measure the threat detection rates for their own patron populations and threat mixtures.

**Ground truth throughput.** Similarly, the throughput speed may not be limited by how fast the screening hardware can process people. It will depend a lot on the number of false alarms, and may be limited by the speed at which people take things out of their pockets, and then pick them up again. It is not uncommon to see walk-through screening temporarily held up while someone reassembles keys, phones, and so forth[9]. This calls for guidelines that help individual venues estimate the throughputs, for their own patron populations, without expensive consultants to do "one time" estimates.

**What Randomization Level to Use?** Even with the facts in hand, each event and crowd presents its own problems. Given the importance of maintaining venue security in today's heightened threat environment, we expect that some organizations or researchers will build an easy-to-use tool (for example, a spreadsheet), together with an inexpensive simulation tool. The spreadsheet could be used to "rough cut" a plan for each specific event, such as the fraction of patrons to be assigned to each of the practices, and report the resulting decrease in missed threats, together with the expected average delay. A simulation tool can also show how much random variation might occur. That way a venue will be able to validate a plan by confirming that the recommended mix will, with high probability[10], generate acceptable throughput.

Finally, the use of randomly mixed practices has some added benefit. Overall threat prevention is the result of both detection and deterrence. The expert responses suggest that a highly visible "high sensitivity" practice, particularly when it is used randomly, will likely deter some threats. As a result, the number of threats that get past security is reduced even more than the mathematical analysis suggests.

---

[9] The reader will notice that if divestment and re-collection are the limiting factors, rather than false alarms, it may be possible to raise the sensitivity without increasing delays.

[10] We need these "weasel words" because both the throughput and the detection process have some randomness themselves, and this means that repeating a simulation will give a somewhat different answer each time.

## 7. Technologies for More Effective Randomization

For any algorithm to be certified by a major agency, such as the TSA, it must undergo expensive and extensive real-world testing to verify its detection capabilities. With current regulations[11], any new specific combination of algorithms must be tested. As one expert noted, these government testing rules, which are quite rigorous, do not make it easy to study combinations of settings. Each possible combination would look like a "new algorithm" from a regulatory standpoint.

> Standards are often years out of date, and are always the result of some negotiations to find a compromise between the highest imaginable security and what is technically feasible.
>
> Many security device engineers.

Along with the human factor, modern security usually combines both a physical device, and a computational or data device. Both chemical and electromagnetic security devices have physical components (sensors) and computational components (algorithms). Every "setting" of a security device represents a choice of the computer algorithm that processes the electronic data coming from the magnetic coils or other sensors within the device.

In principle it seems that those electronic data could be stored and reused to test the effectiveness of alternative algorithms against *exactly the same data stream.* Instead, current testing processes for e.g., bag screening repeat the movement and scanning of large quantities of baggage each time a new algorithm is to be tested. Therefore, validating methods that involve random combination of the data processing algorithms becomes prohibitively expensive

> From the engineering point of view, the laboratory could store every bit of data generated during the test of one algorithm, and present exactly those same cases to a new algorithm, at much lower cost.
>
> Security device engineer.

Instead the signals from the sensors could be captured exactly, and stored by the government's testing laboratories, then the performance of a new algorithm, or combination of algorithms could be tested, using the stored data, at much lower cost than running a complete physical evaluation.

Currently, some security devices do have an "external data port" that can output some, but not all of the data that are fed to the algorithm. However, this "exposed data" is generally not sufficient to compare algorithms. And device manufacturers are unlikely to move, on their own, to reveal the complete data set, for several reasons.

---

[11] Approved lists of systems and devices are issued by the TSA periodically. For example: https://www.tsa.gov/sites/default/files/non-ssi_acstl.pdf  The detailed specifications certified by the testing are Department of Transportation Sensitive Security Information (SSI) and not publicly available.

> Change will only happen if the market demands it, not if a *g*overnmental agency without regulatory authority requests it. Of course, if a large agency, such as the TSA, puts a requirement in their procurement documentation that will also be a motivator.
>
> Test and evaluation expert

First, revealing all the data is more costly. Second, revealing those data might permit reverse-engineering of the sensor technology itself, with possible piracy or copying of proprietary concepts. And, finally, exposing sensor data would open the market to competitors who might sell just the algorithms that are built to work with the output of devices made by other vendors. This would negate the commercial benefits of bundling hardware and software.

With so many commercial risks, there are serious barriers to storage and reuse of sensor outputs. Nonetheless, vendors would provide a full "data readout" if a major agency such as the TSA were to specify that capability for certification. Any change in such regulations will take considerable time, and requires negotiation between government and industry, to reach the best possible outcome with currently available technology. Even so, the long-term benefits to security at venues of every kind from algorithm mixing and frequent changes may justify the effort. One possibility that could be explored is a specification that permits vendors to encrypt their sensor data, so that only their own algorithms can decrypt and read their own stored data.

## 8. Conclusions and Discussion

The study of literature and survey of security experts has revealed that randomization has potential to increase venue security. It also shows that security directors have several different ways of interpreting randomization, and that it serves a number of goals. However, practices today do not fully exploit randomization's potential. Practices are limited by worker capability, and the need to keep job requirements clear and well defined. We have discussed two ways that randomization could be "baked into" security processes without making front line jobs more complex. The first is to use mixed settings on security devices, so that some random fraction of the patrons are screened more thoroughly. We find that this could be done, but *security directors need tools that let them assess detection rates, throughputs, and the random variability of such processes*. The second approach would have mixed settings built into the security devices (sniffers, metal detectors, etc.) themselves. This is within the reach of current engineering, but currently faces very costly validation procedures. We also suggest that government and industry might move towards a regime in which *vendors are permitted to store (encrypted) proprietary physical data at a government testing laboratory, so that new algorithms can be validated against existing data,* at a much lower cost.

## 9. To Read More

The benefits of randomness are described in the Game Theory literature. Some researchers have studied complex situations in which our scoring is different from the adversary's scoring. Those are called "non-zero sum games." Homeland Security researchers at the CREATE Center based at the University of Southern California have developed strategies in which the defender's policy must be random, and it also must be *announced*. It is chosen to drive the adversary into a particular strategy of his own (Kiekintveld et al., 2011, 2009). A remarkably readable explanation of this approach, known as "leader games" or "Stackelberg games" was given in a report by the DARPA think-tank group called JASON on the problem of "Rare Events" and their prediction (JASON, 2009) The explanation begins on page 70 of that report. Unfortunately, there is very little information available about how the adversaries do "score" the possible outcomes, making it hard to apply these concepts in practice.

Another deep discussion of randomness is given by (Press, 2009) His analysis asserts that uniform randomization is better than some kinds of threat-weighted randomness, because, under threat-weighting, we re-examine the same people more often. However, for venue security this might not be relevant, because any given person could become a threat at any time. In recent years there has been little published on random strategies for physical security. One notable exception is (Che et al., 2024).

**Acknowledgments.** This research was supported in part by DHS under Task Order HSHQDC-17-J-00302 – BOA HSHQDC-16-A-B0005 to Rutgers University and NSF award 0735910 to Rutgers University. The authors acknowledge many helpful conversations with Dennis E. Egan, Christie G. Nelson and James Wojtowicz. The survey was developed by the entire CCICADA team, and the survey software was built by Katie McKeon. The literature review was assisted by Jonathan Bullinger. Most importantly, this work is informed by the many security experts and engineers who offered their expertise and insights, but who wish to remain anonymous.

**Disclaimer:** Any ideas, suggestions or conclusions in this paper are those of the authors, and do not represent positions of the Department of Homeland Security, the DHS Office of Safety Act Implementation, the National Science Foundation, the CCICADA Center, or any other institution or organization.